\documentclass[aps,prd,twocolumn,preprintnumbers,amsmath,amssymb,superscriptaddress,floatfix,showpacs,10pt]{revtex4-1}

\usepackage{color}
\usepackage[hyperindex,breaklinks]{hyperref}
\usepackage{graphicx}

\begin{document}

\preprint{DESY 12-131}

\title{Time Evolution of the Large-Scale Tail of Nonhelical Primordial Magnetic Fields with Back-Reaction of the Turbulent Medium}

\author{Andrey Saveliev}
\email{andrey.saveliev@desy.de}
\affiliation{{II}. Institut f\"ur Theoretische Physik, Universit\"at Hamburg, Luruper Chaussee 149, 22761 Hamburg, Germany}
\author{Karsten Jedamzik}
\email{karsten.jedamzik@um2.fr}
\affiliation{Laboratoire de Physique Theorique et Astroparticules, UMR5207-CNRS, Universit\'e Montpellier II, 34095 Montpellier, France}
\author{G\"unter Sigl}
\email{guenter.sigl@desy.de}
\affiliation{{II}. Institut f\"ur Theoretische Physik, Universit\"at Hamburg, Luruper Chaussee 149, 22761 Hamburg, Germany}

\begin{abstract}
We present a derivation of the time evolution equations for the energy content of nonhelical magnetic fields and the accompanying turbulent flows from 
first principles of incompressible magnetohydrodynamics in the general framework of homogeneous and isotropic turbulence. This is then applied to the
early Universe, i.e., the evolution of primordial magnetic fields. Numerically integrating the equations, we find that most of the energy is concentrated 
at an integral wavenumber scale $k_{I}$ where the turbulence turn over time equals the Hubble time. At larger length scales $L$, i.e., smaller wavenumbers 
$q = 2 \pi / L\ll k_I$, independent of the assumed turbulent flow power spectrum, mode-mode coupling tends to develop a small $q$ magnetic field tail 
with a Batchelor spectrum proportional to the fourth inverse power of $L$ and therefore a scaling for the magnetic field of $B \sim L^{-5/2}$.
\end{abstract}

\doi{10.1103/PhysRevD.86.103010}

\maketitle

\section{Introduction}
The question of the origin and time evolution of primordial magnetic fields in the early Universe is an interesting and so far at best partially resolved 
problem in cosmology. It is possible that strong magnetic fields have been created on small scales in the early Universe by a cosmological process, for 
example at a phase transition \cite{PhysRevD.55.4582} or, less likely, during inflation \cite{PhysRevD.37.2743} (for an overview of possible magnetogenesis 
scenarios see, for example, Ref.~\cite{PhysRep.348.163} or Ref.~\cite{Subramanian:2009fu}). A central question then is whether these magnetic fields could 
evolve with time and transport some of their energy content to large scales to account for the recently claimed detection of intergalactic magnetic fields 
\cite{PhysRevD.80.123012,Neronov02042010}. 

Many attempts to study the evolution of primordial magnetic fields have been performed in the past \cite{Dimopoulos:1996nq,Brandenburg:1996fc,Jedamzik:1996wp,
Subramanian:1997gi,Son:1998my,Christensson:2000sp,Sigl:2002kt,Banerjee:2004df,Campanelli:2007tc,Kahniashvili:2010gp}. Numerical simulations are problematic
as they lack the resolution required to give reliable predictions. A further complication in the study of the evolution of magnetic fields in the early 
Universe is the enormous cosmic expansion between a putative magnetogenesis scenario and the present, such that the smallest errors in extrapolation lead
to large changes in the final prediction.

In this paper we take a semianalytic approach which derives the main time evolution equations from first principles of magnetohydrodynamics (MHD), employing 
some fairly generic assumptions. Our analysis follows a similar procedure as has been already applied to, for example, the solar wind \cite{Grappin82,Grappin83} 
or the galactic dynamo problem \cite{Kulsrud:1992rk}.

Two of us \cite{PhysRevD.83.103005} have recently, for the first time, applied such an approach to the evolution of primordial magnetic fields. Considering 
only the most important large $q$ velocity-magnetic field mode-mode coupling source term to generate magnetic fields on small $q$ taken from Ref.~\cite{Kulsrud:1992rk}, 
and building on the result of Banerjee and Jedamzik \cite{Banerjee:2004df}, it has been established that cosmic expansion seems slow enough to allow for 
the generation of large-scale magnetic fields with a white noise, i.e., $B \sim L^{-3/2}$, or even shallower spectrum, also in the absence of initial magnetic 
and velocity fields on such large scales \cite{remark1}. However, the analysis is far from complete as the role of back-reaction of the turbulent medium onto 
the magnetic fields has not been accounted for. Here we present the full analysis including all source and sink terms for nonhelical fields.

This paper is structured as follows: In Sec.~\ref{sec:TimeHomoIsoMHD} we give the evolution equations for the magnetic and turbulence fields in the general
framework of homogeneous and isotropic noncompressible MHD. An outline of the lengthy derivation of these equations is delegated to Appendix \ref{app:calc}, 
whereas a different form more suitable for numerical integrations is presented in Appendix \ref{app:alternative}. In Sec.~\ref{sec:EarlyUniverse} this is 
then extended to the situation in the early Universe with the results and conclusions from these considerations finally presented in Secs.~\ref{sec:Results} 
and \ref{sec:Conclusions}, respectively.

\section{Time Evolution of the Magnetic and Kinetic Energy Content in Homogeneous Isotropic Magnetohydrodynamics} \label{sec:TimeHomoIsoMHD}
The equations describing the time evolution of the two main observables of magnetohydrodynamics, the velocity field of the turbulent medium $\mathbf{v}$ and the 
magnetic field $\mathbf{B}$, are, for an incompressible fluid (i.e., $\nabla\cdot\mathbf{v}=0$), given by
\begin{equation} \label{DiffB}
\partial_{t} \mathbf{B} = \frac{1}{4 \pi \sigma} \Delta \mathbf{B} + \nabla \times \left( \mathbf{v} \times \mathbf{B} \right) 
\end{equation}
and
\begin{equation} \label{Diffv}
\partial_{t} \mathbf{v} = - \left( \mathbf{v} \cdot \nabla \right) \mathbf{v} + \frac{\left( \nabla \times \mathbf{B} \right) \times \mathbf{B}}{4 \pi \rho} + \mathbf{f}_{v}\,,
\end{equation}
respectively. Here $\sigma$ is the conductivity (which is assumed to be very large in the following) and $\rho$ the mass density of the fluid, while $\mathbf{f}_{v}$ 
is some viscous density force. Note that for most part of the evolution of the early Universe incompressibility is an excellent assumption due to the large speed of 
sound in the relativistic plasma.

Our main interest lies in the average buildup of magnetic and kinetic energy density, respectively, taken over an ensemble of cosmic realizations (denoted by chevrons 
$\left\langle \right\rangle$), i.e.,
\begin{equation} \label{dtMdtU}
\left\langle \partial_{t} M_{q} \right\rangle,~\left\langle \partial_{t} U_{q} \right\rangle
\end{equation}
with $M_{q}$ being the magnetic spectral energy defined through
\begin{equation} \label{MagEn}
\epsilon_{B} = \frac{1}{8\pi V}\int {\rm d^{3}}x\, \mathbf{B}^{2}(\mathbf{x}) = \int \frac{{\rm d^{3}}k}{8\pi}\, |\hat{\mathbf{B}}(\mathbf{k})|^{2} \equiv \rho \int {\rm d}k\, M_{k} 
\end{equation}
and $U_{q}$ being the kinetic spectral energy given by
\begin{equation} \label{KinEn}
\epsilon_{K} = \frac{\rho}{2 V} \int {\rm d^{3}}x\, \mathbf{v}^{2}(\mathbf{x}) = \frac{\rho}{2} \int {\rm d^{3}}k\, |\hat{\mathbf{v}}(\mathbf{k})|^{2} \equiv \rho \int {\rm d}k\, U_{k}\,.
\end{equation}
For both expressions $V$ denotes the volume and we have assumed cosmic homogeneity and isotropy which implies that $M_{k}$ and $U_{k}$ are functions only of the 
magnitude $k$ of the wave vector $\mathbf{k}$. Furthermore, Parseval's Theorem has been used in both cases in order to obtain a $k$ integral where a hat denotes 
the Fourier transform normalized by $V^{\frac{1}{2}}$ (cf.~Appendix \ref{app:calc}). With these assumptions the spectral energy densities are therefore

\begin{equation} \label{Mk}
M_{q} = \frac{q^{2}}{2 \rho} |\hat{\mathbf{B}}(\mathbf{q})|^{2}
\end{equation}
and
\begin{equation} \label{Uk}
U_{q} = 2 \pi q^{2} |\hat{\mathbf{v}}(\mathbf{q})|^{2}\,.
\end{equation}

By performing the calculations which are presented in Appendix \ref{app:calc}, we obtain a very general result for the homogeneous and isotropic case which for 
(\ref{dtMdtU}) are given by
\begin{widetext}
\begin{equation} \label{dtMqfinalKJ}
\begin{split}
\left\langle \partial_{t} M_{q} \right\rangle = \int_{0}^{\infty} { \rm d}k \Bigg\{ \Delta t \int_{0}^{\pi} { \rm d}\theta \Bigg[ &- \frac{1}{2} \frac{q^{2} k^{4}}{k_{1}^{4}} \sin^{3}\theta \left\langle M_{q} \right\rangle \left\langle U_{k_{1}} \right\rangle + \frac{1}{2} \frac{q^{4}}{k_{1}^{4}} \left( q^{2} + k^{2} - q k \cos\theta \right) \sin^{3}\theta \left\langle M_{k}\right\rangle \left\langle U_{k_{1}}\right\rangle \\
&- \frac{1}{4} q^{2} \left(3 - \cos^{2}\theta\right) \sin\theta \left\langle M_{k} \right\rangle \left\langle M_{q} \right\rangle \Bigg] \Bigg\}
\end{split}
\end{equation}
and
\begin{equation} \label{dtUqfinalKJ}
\begin{split}
\left\langle \partial_{t} U_{q} \right\rangle &= \int_{0}^{\infty} { \rm d}k \Bigg\{ \Delta t \int_{0}^{\pi} { \rm d}\theta \Bigg[ \frac{1}{4} \frac{q^{3} k}{k_{1}^{4}} \left( q k \sin^{2}\theta + 2 k_{1}^{2} \cos\theta \right) \sin\theta \left\langle M_{k} \right\rangle \left\langle M_{k_{1}} \right\rangle + \frac{1}{4} \frac{q^{4} k}{k_{1}^{4}} \left( 3 k - q \cos\theta \right) \sin^{3}\theta \left\langle U_{k} \right\rangle \left\langle U_{k_{1}} \right\rangle \\
&+ \frac{1}{4} \frac{q k^{4}}{k_{1}^{4}} \left( -3 q + k \cos\theta \right) \sin^{3}\theta \left\langle U_{q} \right\rangle \left\langle U_{k_{1}} \right\rangle - \frac{1}{2} \frac{k^{4}}{k_{1}^{4}} \left( q^{2} + k^{2} - q k \cos\theta \right) \sin^{3}\theta \left\langle M_{k_{1}} \right\rangle \left\langle U_{q} \right\rangle + \frac{1}{2} k^{2} \sin^{3}\theta \left\langle M_{k} \right\rangle \left\langle U_{q} \right\rangle \Bigg] \Bigg\}
\end{split}
\end{equation}
\end{widetext}
where $q$, $k$ and $k_{1}$ are the magnitudes of the wave vectors $\mathbf{q}$, $\mathbf{k}$ and $\mathbf{k_{1}} = \mathbf{q} - \mathbf{k}$, respectively, 
and $\theta$ is the angle between $\mathbf{q}$ and $\mathbf{k}$, i.e., $\mathbf{q} \cdot \mathbf{k} = q\,k\,{\rm cos\,\theta}$. Equations (\ref{dtMqfinalKJ}) 
and (\ref{dtUqfinalKJ}) are a set of well-defined equations since they ensure conservation of energy, momentum and mass density to the lowest nontrivial order 
in $\Delta t$ \cite{remark2}.

\section{Applications to the Early Universe} \label{sec:EarlyUniverse}
In order to study the time evolution of magnetic fields in the early Universe we include Expansion. For the following considerations it is convenient to introduce 
the scale factor $a$ as the time evolution parameter which in this work is normalized such that it is $a=1$ for the initial conditions, i.e., at the magnetogenesis 
era. The focus here will be on the long radiation dominated period in the early Universe where most of the nontrivial evolution occurs. The derivation of the MHD 
equations in an expanding Universe has, for example, been presented in \cite{Jedamzik:1996wp}. These equations contain redshifting terms proportional to powers of 
the scale factor, properly accounting for, e.g., the physical decrease of magnetic fields, i.e.,~$\mathbf{B}\sim 1/a^2$ due to flux-freezing. These terms are not 
taken into account in Eqs.~(\ref{dtMqfinalKJ}) and (\ref{dtUqfinalKJ}) and have to be included now. However, it is known that with a proper scaling of variables 
(cf.~Appendix B in Ref.~\cite{Banerjee:2004df}),

\begin{equation}
{\rm d}t_{c} \equiv {\rm d}t\, a^{-1}\quad \mathbf{B}_{c}\equiv \mathbf{B}\, a^2 \quad \mathbf{v}_{c} \equiv \mathbf{v}\quad \rho_{c} \equiv \rho a^4\quad k_{c} \equiv k a\,,
\end{equation}
the MHD equations in an expanding radiation-dominated Universe are form-invariant to the corresponding equations in a static background. Our master equation is therefore 
also valid in the early Universe when these \textit{comoving} (denoted above by a superscript "c") variables are used. It is convenient to change the time derivative to a 
scale factor derivative
\begin{equation}
\frac{\partial}{\partial t_{c}} = \frac{1}{2 t_{0} a}\frac{\partial}{\partial \ln a} =\frac{H_{0}}{a}\frac{\partial}{\partial \ln a}
\end{equation}
where $t_{0}$ and $H_{0}$ are cosmic time and Hubble parameter at the initial magnetogenesis period, respectively. It is noted here that due to our definition of the 
Fourier Transform given in Appendix \ref{app:calc}, the comoving Fourier Transforms for $\mathbf{B}_{c}$ and $\mathbf{v}_{c}$ are given by
\begin{equation}
\hat{\mathbf{B}}_{c}\equiv \hat{\mathbf{B}}a^{1/2}\quad 
\hat{\mathbf{v}}_{c}\equiv \hat{\mathbf{v}}a^{-3/2}\, . 
\end{equation}
This implies that $M_q^c\equiv M_qa^{-1}$ and $U_q^c\equiv U_qa^{-1}$, such that $\epsilon_B^c$ and $\epsilon_K^c$ are constant during the expansion of the Universe, 
when dynamical evolution is excluded. In what follows, the superscript
$"c"$ is dropped and all variables are comoving, unless specifically
noted otherwise.

It is now possible to rewrite (\ref{dtMqfinalKJ}) and (\ref{dtUqfinalKJ}) in comoving coordinates as
\begin{eqnarray} \label{dtMqfinalKJk2}
\left\langle \frac{\partial M_{q}}{\partial \ln a} \right\rangle 
& = & \frac{a}{H_{0}} \int {\rm d}k \Bigg\{ \Delta t \int{\rm d}\theta \Bigg[ ... \Bigg] \Bigg\} \\
\left\langle \frac{\partial U_{q}}{\partial \ln a} \right\rangle 
\label{dtUqfinalKJk2}
& = & \frac{a}{H_{0}} \int {\rm d}k \Bigg\{ \Delta t \int{\rm d}\theta \Bigg[ ... \Bigg] \Bigg\}\, , 
\end{eqnarray}
where the square brackets $[...]$ in (\ref{dtMqfinalKJk2}) and (\ref{dtUqfinalKJk2}) denote remaining terms in (\ref{dtMqfinalKJ}) and (\ref{dtUqfinalKJ}), respectively.

Finally, a choice for $\Delta t$ has to be made. Following methods applied in molecular chaos we choose $\Delta t$ to be the smallest of all \emph{eddy} turnover 
times, i.e.,~$\Delta t \simeq  {\rm min}(L/v^L,L/v^L_A)$ where $v^L$ and $v^L_A$ are the effective fluid and Alfv\'en velocities on the scale $L$. In particular 
we take
\begin{equation} \label{Dt}
\Delta t \simeq \min\left[\frac{a}{H_{0}},
\frac{2 \pi}{k \left( 2 k \left\langle U_{k} \right\rangle \right)^{\frac{1}{2}}} ,
\frac{2 \pi}{k \left( 3/2 k \left\langle M_{k} \right\rangle\right)^{\frac{1}{2}}}\right]\, ,
\end{equation}
such that we identify $\left( 2 k \left\langle U_{k} \right\rangle \right)^{\frac{1}{2}}$ and $\left( 3/2 k \left\langle M_{k} \right\rangle\right)^{\frac{1}{2}}$ with the effective fluid and Alfv\'en velocities on scale
$L = 2\pi /k$. Note that the first timescale in the brackets of Eq.~(\ref{Dt}) is supposed to ensure causality, in practice, however, our results are independent 
of this condition.

\section{Results} \label{sec:Results}
\begin{figure}
\centering
  \includegraphics[scale=0.306]{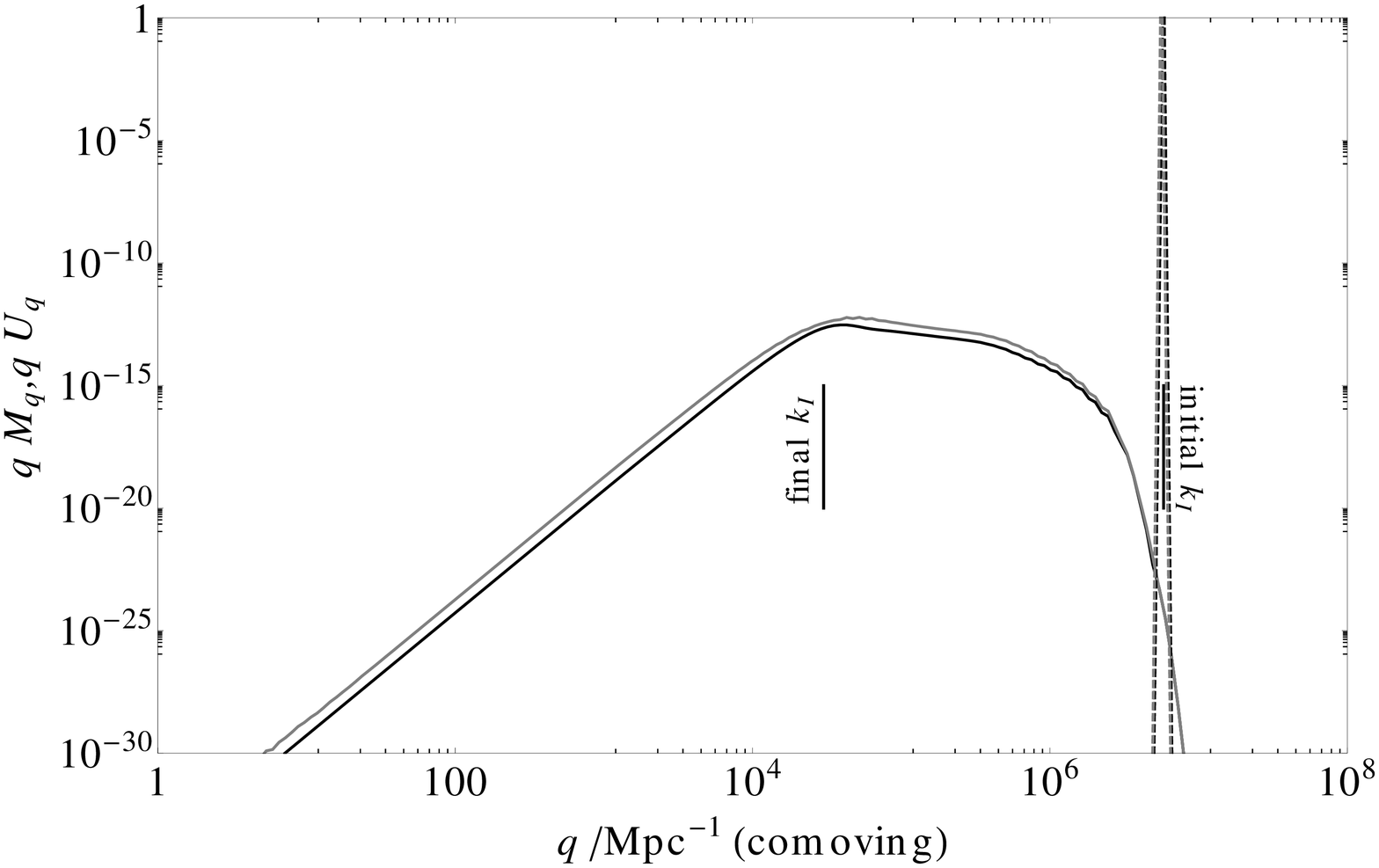}
  \includegraphics[scale=0.306]{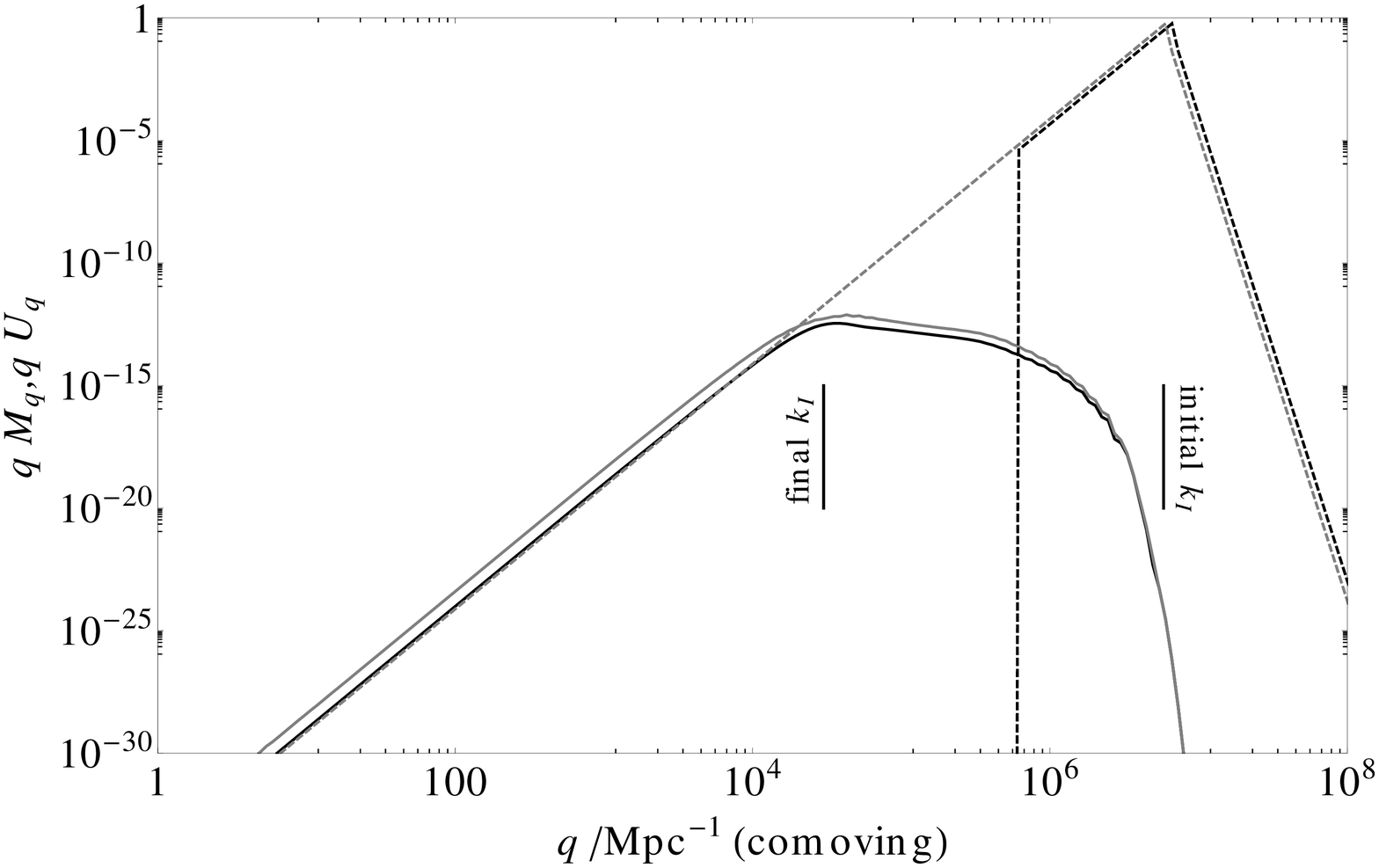}
  \caption{Time evolution of magnetic (black) and kinetic (gray) spectral energies according to (\ref{dtMqfinalKJk2}) and (\ref{dtUqfinalKJk2}), respectively. Dashed 
           lines denote the initial conditions (i.e., at $a=1$) while the solid lines represent the situation for $a=10^{8}$. \emph{Upper panel}: Starting at some 
           time where both spectral energies were concentrated on the same scale (which, in Ref.~\cite{PhysRevD.55.4582}, has been shown to be a reasonable assumption) 
           they evolve close to equipartition, building up a $q^{4}$ slope \emph{ab initio}. \emph{Lower panel}: Starting at some time when the turbulence has already 
           built up a $q^{4}$ slope, after some time the magnetic spectral energy reaches equipartition with the same slope fairly well. It should be noted that for both 
           cases the actual value for $k_{I}$ from the simulation is in good agreement with the one predicted by (\ref{kint}), denoted by the vertical lines labeled 
           ``initial $k_{I}$'' and ``final $k_{I}$'', respectively.}
  \label{fig:timeevolution}
\end{figure}

With this at hand it is now possible to analyze the time development of both the spectral magnetic and kinetic energy including back-reaction by numerically integrating
Eqs.~(\ref{dtMqfinalKJk2}) and (\ref{dtUqfinalKJk2}). We assume two different initial conditions. In all cases it is assumed that magnetic fields and turbulent velocities 
are created on some small scale, called the initial integral scale $k_{I}^{0} \equiv k_{0}$, i.e.,~for $q=k_{0}$ both $\left\langle M_{q} \right\rangle$ and $\left\langle U_{q} \right\rangle$ 
have a sharp peak. The two cases distinguish themselves in the initial conditions for the small-$q$ tail of the turbulent flows, a tail $qU_q\sim q^5$ and no $U_{q}$ 
tail at all. Results of the evolution of $M_q$ and $U_q$ for these initial conditions are shown in Fig.~\ref{fig:timeevolution}.

It is striking to note that (a) the peak and the integral scale follow the analytical prediction very well, (b) the approximate equipartition between magnetic and 
turbulent energy on all scales is achieved and (c) the final result is independent of the initial conditions, predicting a $q M_{q} \sim q^5$ large scale tail for 
the magnetic fields. Here the last result holds even for the case where a $qU_q\sim q^3$, i.e., an initial white noise turbulent spectrum is assumed. Thus the 
conclusion concerning the small $q$ magnetic tail seems very robust. Below analytical arguments are given to support all these results.
\subsection{Evolution of the Integral Scale}
The simulations follow very well the proposed evolution of the peak (i.e., \emph{integral}) scale $k_{I}(a)$ with scale factor $a$. It has been shown \cite{Banerjee:2004df,PhysRevD.83.103005} 
that if the spectral energy has a slope proportional to $q^{\alpha-1}$ ($\alpha > 1$) for $q<k_{I}$, i.e.,
\begin{equation} \label{qSlope}
\left\langle E_{q} \right\rangle \simeq E_{0} \left(\frac{q}{k_{0}}\right)^{\alpha-1}\,,
\end{equation}
where $E$ can be either $M$ or $U$, and $E_{0}$ (i.e., $M_{0}$ or $U_{0}$) is the normalization constant given by the corresponding initial value at $k_{0}$, we have
\begin{equation} \label{kint}
k_{I} \simeq k_{0} a^{-\frac{2}{\alpha+2}}
\end{equation}
for the time dependent integral scale, $k_{0}$ being determined by \cite{PhysRevD.83.103005}
\begin{equation} \label{kint0}
k_{0} \simeq \frac{2 \pi H_{0}}{v_{0}} = \left( \frac{2 \pi^{2} H_{0}^{2}}{U_{0}} \right)^{\frac{1}{3}}\,,
\end{equation}
and
\begin{equation} \label{Eint}
E_{I} \simeq  E_{0} a^{-2 \frac{\alpha-1}{\alpha+2}}
\end{equation}
for the value of the corresponding spectral energy at $k_{I}$. Here, for initial equipartition with $v_{0} \simeq 1$, it has been assumed that $E_{0} = M_{0} = U_{0} = (2 k_{0})^{-1}$.
\subsection{Large-Scale Magnetic Tail}
Large-scale (i.e., small $q$) tails are best discussed in terms of Eqs.~(\ref{dtMqfinalKJk1}) and (\ref{dtUqfinalKJk1}). For large scales, i.e., $q \ll k_{I} \simeq k$, 
we can assume $\left\langle U_{q} \right\rangle \ll \left\langle U_{k} \right\rangle$ and $\left\langle M_{q} \right\rangle \ll \left\langle M_{k} \right\rangle$ and 
therefore neglect the terms containing $\left\langle M_{q} \right\rangle$ and $\left\langle U_{q} \right\rangle$ in (\ref{dtMqfinalKJk1}) and (\ref{dtUqfinalKJk1}), 
leaving the $\left\langle M_{k} \right\rangle \left\langle U_{k_{1}} \right\rangle$ term in the former and both the $\left\langle M_{k} \right\rangle \left\langle M_{k_{1}} \right\rangle$ 
and $\left\langle U_{k} \right\rangle \left\langle U_{k_{1}} \right\rangle$ terms in the latter one.

If we now use the power-law assumption motivated above, i.e., $\left\langle M_{k} \right\rangle \sim k^{\alpha-1}$ and $\left\langle U_{k} \right\rangle \sim k^{\beta-1}$,
it is possible to evaluate the $k_{1}$ integrals which, performing a Taylor series in $q/k$, give
\begin{equation} \label{dtMqfinalKJk4}
\left\langle \frac{\partial M_{q}}{\partial \ln a} \right\rangle \simeq
q^{4} \frac{a}{H_0} \int {\rm d}k \frac{\Delta t}{k^{2}} \left[ \frac{2}{3} \left\langle M_{k} \right\rangle \left\langle U_{k} \right\rangle + \mathcal{O}\left(\left(\frac{q}{k}\right)^{2}\right) \right]
\end{equation}
and
\begin{equation} \label{dtUqfinalKJk4}
\begin{split}
\left\langle \frac{\partial U_{q}}{\partial \ln a} \right\rangle \simeq  q^{4} \frac{a}{H_0} & \int{\rm d}k \frac{\Delta t}{k^{2}} \Bigg[\left\langle U_{k} \right\rangle \left\langle U_{k} \right\rangle\\
&- \frac{\alpha-4}{3} \left\langle M_{k} \right\rangle \left\langle M_{k} \right\rangle + \mathcal{O}\left(\left(\frac{q}{k}\right)^{2}\right) \Bigg]\,.
\end{split}
\end{equation}

One of the main results from these considerations is the fact that for $q\ll k_{I}$ both the magnetic and the kinetic spectral energies have a steep spectrum which is 
proportional to $q^{4}$ or $L^{-4}$, i.e., $\alpha=\beta=5$, where $L$ is the scale of the corresponding field (in this context a power-law spectrum with this specific 
slope is also called a Batchelor or von K\'arm\'an spectrum). This would correspond to a behavior of the form $B \sim L^{-\frac{5}{2}}$ and $v \sim L^{-\frac{5}{2}}$ 
for the scale dependence of the magnetic and the turbulence fields on large scales, respectively. For the former this would be in good agreement with the analytical 
considerations in Refs.~\cite{Batchelor1956,Durrer:2003ja}, as well as with the numerical results in Ref.~\cite{Tevzadze:2012kk}, which predict the same scaling. For 
the turbulent field it is an important result which is also confirmed by simulation as it can be seen in Fig.~\ref{fig:timeevolution}: fairly independent of the initial
conditions, even starting off with the total energy being concentrated on only one scale, a $q^{4}$ slope for the spectral energies forms.

It is important to estimate the scale factor $a$ at which the Batchelor spectrum can be built up. To do so, as initial conditions we assume a sharp peak with power-law 
slopes on both sides of the peak, for both $U$ and $M$, i.e., for $a=1$ we have
\begin{equation} \label{Powerlawpeak}
\langle M_{k} \rangle = \langle U_{k} \rangle \equiv \langle E_{k} \rangle = \begin{cases}&E_{0} \left( \frac{k}{k_{0}} \right)^{\gamma-1},~k \leq k_{0}\\&E_{0} \left( \frac{k}{k_{0}} \right)^{-(\gamma-1)},~k > k_{0}\,,\end{cases}
\end{equation}
the exponent $\gamma > 5$ determining the width of the peak. Assuming furthermore that at $a \simeq 1$ for (\ref{Dt}) we have $\Delta t = a/H_{0}$, (\ref{dtMqfinalKJk4}) 
becomes
\begin{equation} \label{ApproxUM}
\left\langle \frac{\partial M_{q}}{\partial a} \right\rangle \simeq \frac{2}{3} q^{4} \frac{a}{H_{0}^{2}} \int \frac{{\rm d}k}{k^{2}} \left\langle E_{k} \right\rangle^2 \stackrel{(\ref{Powerlawpeak})}{\simeq} \frac{2}{3} \frac{a}{H_{0}^{2}} \frac{ q^{4} E_{0}^{2}}{(\gamma-1) k_{0}}\,.
\end{equation}
Integrating this equation at $q \ll k_{0}$ from $a=1$, when $M_{q} \simeq 0$, to $a_{LS}$, the scale factor at which the final configuration, i.e., (\ref{qSlope}), is 
approximately established, gives
\begin{equation}
\int_{0}^{E_{0} \left(\frac{q}{k_{0}}\right)^{4}}{\rm d} \left\langle M_{q} \right\rangle \simeq \frac{2}{3} \frac{1}{H_{0}^{2}} \frac{ q^{4} E_{0}^{2}}{(\gamma-1) k_{0}}\ \int_{1}^{a_{LS}} a {\rm d}a\,.
\end{equation}
Using $E_{0} = (2 k_{0})^{-1}$ and $H_{0} = k_{0}/(2 \pi)$ as stated above, this gives the upper limit
\begin{equation}
a_{LS} \lesssim \left( \frac{3}{2 \pi^{2}}  (\gamma-1) + 1 \right)^{\frac{1}{2}} \,.
\end{equation}
A relative full width at half maximum of $\Delta_{\frac{1}{2}}/k_{0} = 1/100$ corresonds to $\gamma \simeq 15$ and gives $a_{LS} \lesssim 5$ which implies that the $q^{4}$ slope builds up after a rather short time.

\subsection{Equipartition}
Furthermore, the system tends to achieve equipartition quite accurately. A simple argument for this at large scales can be extracted from (\ref{dtMqfinalKJk4}) and 
(\ref{dtUqfinalKJk4}): Assuming that the magnetic and kinetic spectral energies have similar values at $k$, i.e., $\left\langle M_{k} \right\rangle \simeq \left\langle U_{k} \right\rangle$, 
and $\alpha=5$ as discussed before, we obtain
\begin{equation}
\left\langle \frac{\partial U_{q}}{\partial \ln a} \right\rangle \simeq \left\langle \frac{\partial M_{q}}{\partial \ln a} \right\rangle \simeq q^{4} \frac{a}{H_{0}} \int {\rm d}k \left( \frac{2 \Delta t}{3 k^{2}} \left\langle U_{k} \right\rangle \left\langle U_{k} \right\rangle \right)\,,
\end{equation}
which means that the time evolution of both is the same to the fourth order in $q$. 

This is a crucial result since it means that equipartition is indeed possible or at least that at some point, even for the case that $M_{q}$ and $U_{q}$ do not have the same magnitude, 
the time development of both at large scales will qualitatively be the same.
\subsection{Estimate of Present Magnetic Field Strength}
To estimate the magnetic field strength $B(L)$ at some scale $L=2 \pi / q$ we use
\begin{equation}
M_{q} = \frac{B^{2}(L)}{8 \pi q \rho}
\end{equation}
and, with (\ref{kint})-(\ref{Eint}), obtain
\begin{equation}
B(L) = \left(8 \pi q \rho M_{q} \right)^{\frac{1}{2}} = B_{0} \left(2 q M_{q} \right)^{\frac{1}{2}} \simeq B_{0} \left( H_{0} L \right)^{-\frac{\alpha}{2}}
\end{equation}
where $B_{0} = (4 \pi \rho)^{\frac{1}{2}}$ is the effective magnetic field for $v_{0} \simeq 1$, i.e., the speed of light, for the initial equipartition of radiation 
and magnetic energies which would correspond to $B_{0} \simeq 3 \times 10^{-6}\,{\rm G}$. If we now acknowledge the integral scale to be the coherence scale of the 
magnetic field, then for the magnetic field strength at $L_{I} = 2 \pi / k_{I}$ we get, using (\ref{kint}) and (\ref{Eint}),
\begin{equation}
B(L_{I}) = B_{0} \left(2 k_{I} M_{I} \right)^{\frac{1}{2}} = B_{0} a^{ \frac{\alpha}{\alpha+2}}
\end{equation}
and therefore for $a=10^{8}$ and $\alpha=5$ it is $L_{I} \simeq 200\,{\rm pc}$ and hence $B(200~{\rm pc}) \lesssim 5 \times 10^{-12}\,{\rm G}$ for the QCD phase 
transition. It is hard to imagine how causally generated magnetic fields in the early Universe could yield a larger remnant field than quoted above, unless they are 
generated with substantial helicity.

\section{Conclusions} \label{sec:Conclusions}
Starting from first principles and reasonable assumptions about homogeneous and isotropic magneto-hydrodynamic turbulence, evolution equations
for the spectral magnetic and kinetic energy densities have been derived. After adapting these equations to the use in an expanding Universe they
were numerically integrated for a number of assumed initial conditions, as possibly resulting during an early magnetogenesis period. It has been found
that, seemingly independent of initial conditions, a $B(L)\sim L^{-5/2}$ tail on scales $L \gg 2 \pi/k_{I}$ always develops. This magnetic field spectrum 
may therefore be regarded as the natural one for nonhelical cosmic magnetic fields (cf. also to Ref.~\cite{Durrer:2003ja}). At the same time most of the energy 
is concentrated on the integral scale which allows a rather generic prediction of coherence scale and strength of causally created primordial magnetic fields.

\begin{acknowledgments}
This work was supported by the Deutsche Forschungsgemeinschaft through the collaborative research centre SFB 676, by the ``Helmholtz Alliance for
Astroparticle Phyics (HAP)'' funded by the Initiative and Networking Fund of the Helmholtz Association, and by the State of Hamburg, through the
Collaborative Research program ``Connecting Particles with the Cosmos''.
\end{acknowledgments}

\appendix
\section{Derivation of the Master Equations (\ref{dtMqfinalKJ}) and (\ref{dtUqfinalKJ})} \label{app:calc}
The first step in order to derive (\ref{dtMqfinalKJ}) and (\ref{dtUqfinalKJ}) is to obtain an expression for (\ref{Mk}) and (\ref{Uk}). To do so we first 
take the Fourier transforms of the initial equations (\ref{DiffB}) and (\ref{Diffv}), where we define the Fourier transform of a vector field $\mathbf{A}(\mathbf{x})$ by
\begin{equation}
{\mathbf{A}(\mathbf{x})} = \frac{V^{\frac{1}{2}}}{(2\pi)^{\frac{3}{2}}} \int {{\rm d^{3}}k}\, \hat{\mathbf{{A}}}(\mathbf{k})\, {\rm e}^{i\mathbf{k} \cdot \mathbf{x}} \,.
\end{equation}
Using the convolution theorem, for a wave vector $\mathbf{q}$ they are given by
\begin{equation} \label{DiffBFT1}
\begin{split}
\partial_{t} \hat{\mathbf{B}}(\mathbf{q}) &= - \frac{1}{4 \pi \sigma} q^{2} \hat{\mathbf{B}}(\mathbf{q}) \\
&+ \frac{i V^{\frac{1}{2}}}{(2 \pi)^{\frac{3}{2}}} \int {\rm d^{3}}k \left\{ \mathbf{q} \times \left[ \hat{\mathbf{v}}(\mathbf{q} - \mathbf{k}) \times \hat{\mathbf{B}}(\mathbf{k}) \right] \right\}
\end{split}
\end{equation}
and
\begin{equation} \label{Diffvt}
\begin{split}
\partial_{t} \hat{\mathbf{v}}(\mathbf{q}) &= - \frac{i V^{\frac{1}{2}}}{(2 \pi)^{\frac{3}{2}}} \int {\rm d^{3}}k \left\{ \Big[ \mathbf{k} \cdot \hat{\mathbf{v}}(\mathbf{q} - \mathbf{k}) \Big] \hat{\mathbf{v}}(\mathbf{k}) \right\} \\
&+ \frac{i V^{\frac{1}{2}}}{(2 \pi)^{\frac{3}{2}}} \frac{1}{4 \pi \rho} \int d^{3} k \left\{ \left[\mathbf{k} \times \hat{\mathbf{B}}(\mathbf{k}) \right] \times \hat{\mathbf{B}}(\mathbf{q} - \mathbf{k}) \right\}.
\end{split}
\end{equation}
We solve these ordinary differential equations by using the midpoint method which gives us $\hat{\mathbf{B}}(\mathbf{q},t)$ and $\hat{\mathbf{v}}(\mathbf{q},t)$ 
at the time $t = t_{0}+\Delta t$ for some initial conditions $\hat{\mathbf{B}}_{0}(\mathbf{q}) = \hat{\mathbf{B}}(\mathbf{q},t_{0})$ and $\hat{\mathbf{v}}_{0}(\mathbf{q}) = \hat{\mathbf{v}}(\mathbf{q},t_{0})$.
In particular, if the time derivative of some function $A$ is $\partial_t A = f(t,A(t))$, the midpoint method is given by
\begin{equation}
A(t + \Delta t) \! = \! A(t) + \Delta t f\bigg(t + \Delta t/2 ,A(t) + \Delta t/2\, f\Big[t,A(t)\Big]\bigg)\, .
\end{equation}
We use this method as it is accurate to second order in the time step $\Delta t$, essential for energy conservation of Eqs.~(\ref{dtMqfinalKJ}) and (\ref{dtUqfinalKJ}).

In order to calculate (\ref{dtMdtU}) we use (\ref{Mk}) and (\ref{Uk}) and therefore obtain
\begin{equation} \label{Mqt}
\begin{split}
\allowdisplaybreaks
&\left\langle \partial_{t} M_{q} \right\rangle \\
&= \left\langle \partial_{t} \left[ \frac{q^{2}}{2 \rho} |\hat{\mathbf{B}}(\mathbf{q})|^{2} \right] \right\rangle =  \left\langle \frac{q^{2}}{2\rho} \left[ \partial_{t}\hat{\mathbf{B}}(\mathbf{q}) \right] \cdot \hat{\mathbf{B}}(\mathbf{q})^{*} \right\rangle +{\rm H.c.}\\
&\simeq \left\langle \frac{q^2}{2\rho} \frac{\hat{\mathbf{B}}(\mathbf{q},t) \cdot \hat{\mathbf{B}}(\mathbf{q},t)^* - \hat{\mathbf{B}}(\mathbf{q},t_{0}) \cdot \hat{\mathbf{B}}(\mathbf{q},t_{0})^*}{\Delta t}\right\rangle
\end{split}
\end{equation}
and

\begin{equation} \label{Uqt}
\begin{split}
&\left\langle \partial_{t} U_{q} \right\rangle \\
&= \left\langle \partial_{t} \left[ 2 \pi q^{2} |\hat{\mathbf{v}}(\mathbf{q})|^{2} \right] \right\rangle = \left\langle 2 \pi q^{2} \Big[ \partial_{t}\hat{\mathbf{v}}(\mathbf{q}) \Big] \cdot \hat{\mathbf{v}}(\mathbf{q})^{*} \right\rangle + {\rm H.c.}\\
&\simeq \left\langle 2 \pi q^2 \frac{\hat{\mathbf{v}}(\mathbf{q},t) \cdot \hat{\mathbf{v}}(\mathbf{q},t)^* - \hat{\mathbf{v}}(\mathbf{q},t_{0}) \cdot \hat{\mathbf{v}}(\mathbf{q},t_{0})^*}{\Delta t}\right\rangle
\end{split}
\end{equation}
which are accurate to order $\Delta t$ if the midpoint method is used.

When the terms $\hat{\mathbf{B}}(\mathbf{q},t) \cdot \hat{\mathbf{B}}(\mathbf{q},t)^*$ and $\hat{\mathbf{v}}(\mathbf{q},t) \cdot \hat{\mathbf{v}}(\mathbf{q},t)^*$
with $t = t_{0} + \Delta t$ in Eqs.~(\ref{Mqt}) and (\ref{Uqt}) are evaluated with the help of Eqs.~(\ref{DiffBFT1}) and (\ref{Diffvt}) the result is a large number of
expressions of a similar structure such as 

\begin{equation} \label{example}
\begin{split}
\sim &(\Delta t)^2 \biggl\langle q^{2} \int{\rm d^{3}}k \int{\rm d^{3}}k^{\prime} \biggl[ \biggl( \mathbf{q} \cdot {\hat{\mathbf{B}}}(\mathbf{k},t)\biggr) \biggl( \mathbf{q} \cdot \hat{\mathbf{B}}(\mathbf{k^{\prime}},t)\biggr) \\
&\times \biggl(\hat{\mathbf{v}}(\mathbf{q-k},t) \cdot \hat{\mathbf{v}}(\mathbf{q-k^{\prime}},t)\biggr) \biggl] \biggl\rangle\,.
\end{split}
\end{equation}

Making the reasonable assumption of approximately uncorrelated statistical chaos in homogeneous and isotropic nonhelical MHD 
turbulence, one may write (cf. Ref.~\cite{ProcRSocLondonA.164.192})
\begin{eqnarray} \label{corr1}
\left\langle \hat{B}_{a}(\mathbf{k},t) \hat{B}_{b}(\mathbf{k'},t')^{*} \right\rangle &\simeq& C_{1} \delta_{\mathbf{k}\mathbf{k'}} \delta_{tt'} M_{k}\left( \delta_{ab} - \frac{k_{a} k_{b}}{k^{2}} \right) \label{ensrules1} \\
\label{corr2} 
\left\langle \hat{v}_{a}(\mathbf{k},t) \hat{v}_{b}(\mathbf{k'},t')^{*} \right\rangle &\simeq& C_{2} \delta_{\mathbf{k}\mathbf{k'}} \delta_{tt'} U_{k}\left( \delta_{ab} - \frac{k_{a} k_{b}}{k^{2}} \right) \label{ensrules2} \\
\label{corr3} 
\left\langle \hat{B}_{a}(\mathbf{k},t) \hat{v}_{b}(\mathbf{k'},t')^{*} \right\rangle &\simeq& 0\,, \label{ensrules3}
\end{eqnarray}
where the $\delta$'s denote delta functions and $C_{1}$ and $C_{2}$ are constants to be determined. Using Wick's Theorem for Gaussian fields, i.e.,
\begin{equation}
\left\langle A_{a} A_{b} A_{c} A_{d} \right\rangle = \left\langle A_{a} A_{b} \right\rangle \left\langle A_{c} A_{d} \right\rangle
+\, {\rm other\, permutations}
\end{equation}
one may use the relations (\ref{corr1}) - (\ref{corr3}) to work out one integral in expressions of the type of (\ref{example}), yielding, after computing
many terms of this sort, the final result, Eqs.~(\ref{dtMqfinalKJ}) and (\ref{dtUqfinalKJ}). Finally it is noted that $C_{1} = \rho /k^{2}$ and $C_{2} = \rho/(4\pi k^{2})$ 
as determined from the definitions in Eqs.~(\ref{MagEn}) and (\ref{KinEn}).

\begin{widetext}

\section{Alternative form of the Master Equations} \label{app:alternative}
We give here an alternative formulation of the master equations (\ref{dtMqfinalKJ}) and  (\ref{dtUqfinalKJ}) in terms of $\mathbf{k_{1}} = \mathbf{q} - \mathbf{k}$ 
which is more suitable for numerical integration,

\begin{equation} \label{dtMqfinalKJk1}
\begin{split} 
&\left\langle \partial_{t} M_{q} \right\rangle = \int_{0}^{\infty} {\rm d}k \Bigg\{ \Delta t \int_{|q-k|}^{q+k} {\rm d}k_{1} \Bigg[ \left( \frac{k^{5}}{8 q k_{1}^{3}} - \frac{k^{3}}{4 q k_{1}} - \frac{q k^{3}}{4 k_{1}^{3}} + \frac{k k_{1}}{8 q} - \frac{q k}{4 k_{1}} + \frac{q^{3} k}{8 k_{1}^{3}} \right) \left\langle M_{q} \right\rangle \left\langle U_{k_{1}} \right\rangle \\
& +\left( - \frac{q^{7}}{16 k^{3} k_{1}^{3}} + \frac{q^{5}}{16 k^{3} k_{1}} + \frac{q^{5}}{16 k k_{1}^{3}} + \frac{ q^{3} k}{16 k_{1}^{3}} + \frac{3 q^{3}}{8 k k_{1}} + \frac{q^{3} k_{1}}{16 k^{3}} - \frac{q k^{3}}{16 k_{1}^{3}} + \frac{q k}{16 k_{1}} + \frac{q k_{1}}{16 k} - \frac{q k_{1}^{3}}{16 k^{3}} \right) \left\langle M_{k} \right\rangle \left\langle U_{k_{1}} \right\rangle \\
&+ \left( \frac{k_{1}^{5}}{16 q k^{3}} - \frac{q k_{1}^{3}}{8 k^{3}} - \frac{k_{1}^{3}}{8 q k} + \frac{q^{3} k_{1}}{16 k^{3}} - \frac{5 q k_{1}}{8 k} + \frac{k k_{1}}{16 q} \right) \left\langle M_{q} \right\rangle \left\langle M_{k} \right\rangle \Bigg] \Bigg\}
\end{split}
\end{equation}
and
\begin{equation} \label{dtUqfinalKJk1}
\begin{split}
&\left\langle \partial_{t} U_{q} \right\rangle = \int_{0}^{\infty} {\rm d}k \Bigg\{ \Delta t \int_{|q-k|}^{q+k} { \rm d}k_{1} \Bigg[ \left( - \frac{q^{5}}{16 k k_{1}^{3}} + \frac{q^{3} k}{8 k_{1}^{3}} + \frac{3 q^{3}}{8 k k_{1}} - \frac{q k^{3}}{16 k_{1}^{3}} + \frac{3 q k}{8 k_{1}} - \frac{5 q k_{1}}{16 k} \right) \left\langle M_{k} \right\rangle \left\langle M_{k_{1}} \right\rangle \\
&+ \left( \frac{q^{7}}{32 k^{3} k_{1}^{3}} - \frac{7 q^{5}}{32 k k_{1}^{3}} - \frac{3 q^{5}}{32 k^{3} k_{1}} + \frac{11 q^{3} k}{32 k_{1}^{3}}+ \frac{5 q^{3}}{16 k k_{1}} + \frac{3 q^{3} k_{1}}{32 k^{3}} - \frac{5 q k^{3}}{32 k_{1}^{3}} + \frac{9 q k}{32 k_{1}} - \frac{3 q k_{1}}{32 k} - \frac{q k_{1}^{3}}{32 k^{3}} \right)  \left\langle U_{k} \right\rangle \left\langle U_{k_{1}} \right\rangle \\
&+ \left( - \frac{k^{7}}{32 q^{3} k_{1}^{3}} + \frac{7 k^{5}}{32 q k_{1}^{3}} + \frac{3 k^{5}}{32 q^{3} k_{1}} - \frac{11 q k^{3}}{32 k_{1}^{3}} - \frac{5 k^{3}}{16 q k_{1}} - \frac{3 k^{3} k_{1}}{32 q^{3}} + \frac{5 q^{3} k}{32 k_{1}^{3}} - \frac{9 q k}{32 k_{1}} + \frac{3 k k_{1}}{32 q} + \frac{k k_{1}^{3}}{32 q^{3}} \right) \left\langle U_{q} \right\rangle \left\langle U_{k_{1}} \right\rangle \\
&+ \left( \frac{k^{7}}{16 q^{3} k_{1}^{3}} - \frac{k^{5}}{16 q k_{1}^{3}} - \frac{k^{5}}{16 q^{3} k_{1}} - \frac{q k^{3}}{16 k_{1}^{3}} - \frac{3 k^{3}}{8 q k_{1}} - \frac{k^{3} k_{1}}{16 q^{3}} - \frac{q k}{16 k_{1}} + \frac{q^{3} k}{16 k_{1}^{3}} - \frac{k k_{1}}{16 q} + \frac{k k_{1}^{3}}{16 q^{3}}\right)  \left\langle M_{k_{1}} \right\rangle \left\langle U_{q} \right\rangle \Big\}\\
&+ \left( - \frac{k_{1}^{5}}{8 q^{3} k} + \frac{k_{1}^{3}}{4 q k} + \frac{k k_{1}^{3}}{4 q^{3}} - \frac{q k_{1}}{8 k} + \frac{k k_{1}}{4 q} - \frac{k^{3} k_{1}}{8 q^{3}} \right)\left\langle M_{k} \right\rangle \left\langle U_{q} \right\rangle \Bigg] \Bigg\}\,.
\end{split}
\end{equation}
\end{widetext}


\begin{thebibliography}{26}%
\makeatletter
\providecommand \@ifxundefined [1]{%
 \@ifx{#1\undefined}
}%
\providecommand \@ifnum [1]{%
 \ifnum #1\expandafter \@firstoftwo
 \else \expandafter \@secondoftwo
 \fi
}%
\providecommand \@ifx [1]{%
 \ifx #1\expandafter \@firstoftwo
 \else \expandafter \@secondoftwo
 \fi
}%
\providecommand \natexlab [1]{#1}%
\providecommand \enquote  [1]{``#1''}%
\providecommand \bibnamefont  [1]{#1}%
\providecommand \bibfnamefont [1]{#1}%
\providecommand \citenamefont [1]{#1}%
\providecommand \href@noop [0]{\@secondoftwo}%
\providecommand \href [0]{\begingroup \@sanitize@url \@href}%
\providecommand \@href[1]{\@@startlink{#1}\@@href}%
\providecommand \@@href[1]{\endgroup#1\@@endlink}%
\providecommand \@sanitize@url [0]{\catcode `\\12\catcode `\$12\catcode
  `\&12\catcode `\#12\catcode `\^12\catcode `\_12\catcode `\%12\relax}%
\providecommand \@@startlink[1]{}%
\providecommand \@@endlink[0]{}%
\providecommand \url  [0]{\begingroup\@sanitize@url \@url }%
\providecommand \@url [1]{\endgroup\@href {#1}{\urlprefix }}%
\providecommand \urlprefix  [0]{URL }%
\providecommand \Eprint [0]{\href }%
\providecommand \doibase [0]{http://dx.doi.org/}%
\providecommand \selectlanguage [0]{\@gobble}%
\providecommand \bibinfo  [0]{\@secondoftwo}%
\providecommand \bibfield  [0]{\@secondoftwo}%
\providecommand \translation [1]{[#1]}%
\providecommand \BibitemOpen [0]{}%
\providecommand \bibitemStop [0]{}%
\providecommand \bibitemNoStop [0]{.\EOS\space}%
\providecommand \EOS [0]{\spacefactor3000\relax}%
\providecommand \BibitemShut  [1]{\csname bibitem#1\endcsname}%
\let\auto@bib@innerbib\@empty
\bibitem [{\citenamefont {Sigl}\ \emph {et~al.}(1997)\citenamefont {Sigl},
  \citenamefont {Olinto},\ and\ \citenamefont {Jedamzik}}]{PhysRevD.55.4582}%
  \BibitemOpen
  \bibfield  {author} {\bibinfo {author} {\bibfnamefont {G.}~\bibnamefont
  {Sigl}}, \bibinfo {author} {\bibfnamefont {A.~V.}\ \bibnamefont {Olinto}}, \
  and\ \bibinfo {author} {\bibfnamefont {K.}~\bibnamefont {Jedamzik}},\ }\href
  {\doibase 10.1103/PhysRevD.55.4582} {\bibfield  {journal} {\bibinfo
  {journal} {Phys. Rev. D}\ }\textbf {\bibinfo {volume} {55}},\ \bibinfo
  {pages} {4582} (\bibinfo {year} {1997})},\ \Eprint
  {http://arxiv.org/abs/astro-ph/9610201} {astro-ph/9610201} \BibitemShut
  {NoStop}%
\bibitem [{\citenamefont {Turner}\ and\ \citenamefont
  {Widrow}(1988)}]{PhysRevD.37.2743}%
  \BibitemOpen
  \bibfield  {author} {\bibinfo {author} {\bibfnamefont {M.~S.}\ \bibnamefont
  {Turner}}\ and\ \bibinfo {author} {\bibfnamefont {L.~M.}\ \bibnamefont
  {Widrow}},\ }\href {\doibase 10.1103/PhysRevD.37.2743} {\bibfield  {journal}
  {\bibinfo  {journal} {Phys. Rev. D}\ }\textbf {\bibinfo {volume} {37}},\
  \bibinfo {pages} {2743} (\bibinfo {year} {1988})}\BibitemShut {NoStop}%
\bibitem [{\citenamefont {Grasso}\ and\ \citenamefont
  {Rubinstein}(2001)}]{PhysRep.348.163}%
  \BibitemOpen
  \bibfield  {author} {\bibinfo {author} {\bibfnamefont {D.}~\bibnamefont
  {Grasso}}\ and\ \bibinfo {author} {\bibfnamefont {H.~R.}\ \bibnamefont
  {Rubinstein}},\ }\href
  {http://www.sciencedirect.com/science/article/pii/S0370157300001101}
  {\bibfield  {journal} {\bibinfo  {journal} {Phys. Rep.}\ }\textbf {\bibinfo
  {volume} {348}},\ \bibinfo {pages} {163} (\bibinfo {year} {2001})},\ \Eprint
  {http://arxiv.org/abs/astro-ph/0009061} {arXiv:astro-ph/0009061} \BibitemShut
  {NoStop}%
\bibitem [{\citenamefont {Subramanian}(2010)}]{Subramanian:2009fu}%
  \BibitemOpen
  \bibfield  {author} {\bibinfo {author} {\bibfnamefont {K.}~\bibnamefont
  {Subramanian}},\ }\href {\doibase 10.1002/asna.200911312} {\bibfield
  {journal} {\bibinfo  {journal} {Astron. Nachr.}\ }\textbf {\bibinfo {volume}
  {331}},\ \bibinfo {pages} {110} (\bibinfo {year} {2010})},\ \Eprint
  {http://arxiv.org/abs/0911.4771} {arXiv:0911.4771 [astro-ph.CO]} \BibitemShut
  {NoStop}%
\bibitem [{\citenamefont {Neronov}\ and\ \citenamefont
  {Semikoz}(2009)}]{PhysRevD.80.123012}%
  \BibitemOpen
  \bibfield  {author} {\bibinfo {author} {\bibfnamefont {A.}~\bibnamefont
  {Neronov}}\ and\ \bibinfo {author} {\bibfnamefont {D.~V.}\ \bibnamefont
  {Semikoz}},\ }\href {\doibase 10.1103/PhysRevD.80.123012} {\bibfield
  {journal} {\bibinfo  {journal} {Phys. Rev. D}\ }\textbf {\bibinfo {volume}
  {80}},\ \bibinfo {pages} {123012} (\bibinfo {year} {2009})},\ \Eprint
  {http://arxiv.org/abs/0910.1920} {arXiv:0910.1920 [astro-ph.CO]} \BibitemShut
  {NoStop}%
\bibitem [{\citenamefont {Neronov}\ and\ \citenamefont
  {Vovk}(2010)}]{Neronov02042010}%
  \BibitemOpen
  \bibfield  {author} {\bibinfo {author} {\bibfnamefont {A.}~\bibnamefont
  {Neronov}}\ and\ \bibinfo {author} {\bibfnamefont {I.}~\bibnamefont {Vovk}},\
  }\href {\doibase 10.1126/science.1184192} {\bibfield  {journal} {\bibinfo
  {journal} {Science}\ }\textbf {\bibinfo {volume} {328}},\ \bibinfo {pages}
  {73} (\bibinfo {year} {2010})},\ \Eprint {http://arxiv.org/abs/1006.3504}
  {arXiv:1006.3504 [astro-ph.HE]} \BibitemShut {NoStop}%
\bibitem [{\citenamefont {Dimopoulos}\ and\ \citenamefont
  {Davis}(1997)}]{Dimopoulos:1996nq}%
  \BibitemOpen
  \bibfield  {author} {\bibinfo {author} {\bibfnamefont {K.}~\bibnamefont
  {Dimopoulos}}\ and\ \bibinfo {author} {\bibfnamefont {A.-C.}\ \bibnamefont
  {Davis}},\ }\href
  {http://www.sciencedirect.com/science/article/pii/S0370269396013664}
  {\bibfield  {journal} {\bibinfo  {journal} {Phys. Lett. B}\ }\textbf
  {\bibinfo {volume} {390}},\ \bibinfo {pages} {87} (\bibinfo {year} {1997})},\
  \Eprint {http://arxiv.org/abs/astro-ph/9610013} {arXiv:astro-ph/9610013}
  \BibitemShut {NoStop}%
\bibitem [{\citenamefont {Brandenburg}\ \emph {et~al.}(1996)\citenamefont
  {Brandenburg}, \citenamefont {Enqvist},\ and\ \citenamefont
  {Olesen}}]{Brandenburg:1996fc}%
  \BibitemOpen
  \bibfield  {author} {\bibinfo {author} {\bibfnamefont {A.}~\bibnamefont
  {Brandenburg}}, \bibinfo {author} {\bibfnamefont {K.}~\bibnamefont
  {Enqvist}}, \ and\ \bibinfo {author} {\bibfnamefont {P.}~\bibnamefont
  {Olesen}},\ }\href {\doibase 10.1103/PhysRevD.54.1291} {\bibfield  {journal}
  {\bibinfo  {journal} {Phys. Rev. D}\ }\textbf {\bibinfo {volume} {54}},\
  \bibinfo {pages} {1291} (\bibinfo {year} {1996})},\ \Eprint
  {http://arxiv.org/abs/astro-ph/9602031} {arXiv:astro-ph/9602031} \BibitemShut
  {NoStop}%
\bibitem [{\citenamefont {Jedamzik}\ \emph {et~al.}(1998)\citenamefont
  {Jedamzik}, \citenamefont {Katalinic},\ and\ \citenamefont
  {Olinto}}]{Jedamzik:1996wp}%
  \BibitemOpen
  \bibfield  {author} {\bibinfo {author} {\bibfnamefont {K.}~\bibnamefont
  {Jedamzik}}, \bibinfo {author} {\bibfnamefont {V.}~\bibnamefont {Katalinic}},
  \ and\ \bibinfo {author} {\bibfnamefont {A.~V.}\ \bibnamefont {Olinto}},\
  }\href {\doibase 10.1103/PhysRevD.57.3264} {\bibfield  {journal} {\bibinfo
  {journal} {Phys. Rev. D}\ }\textbf {\bibinfo {volume} {57}},\ \bibinfo
  {pages} {3264} (\bibinfo {year} {1998})},\ \Eprint
  {http://arxiv.org/abs/astro-ph/9606080} {arXiv:astro-ph/9606080} \BibitemShut
  {NoStop}%
\bibitem [{\citenamefont {Subramanian}\ and\ \citenamefont
  {Barrow}(1998)}]{Subramanian:1997gi}%
  \BibitemOpen
  \bibfield  {author} {\bibinfo {author} {\bibfnamefont {K.}~\bibnamefont
  {Subramanian}}\ and\ \bibinfo {author} {\bibfnamefont {J.~D.}\ \bibnamefont
  {Barrow}},\ }\href {\doibase 10.1103/PhysRevD.58.083502} {\bibfield
  {journal} {\bibinfo  {journal} {Phys. Rev. D}\ }\textbf {\bibinfo {volume}
  {58}},\ \bibinfo {pages} {083502} (\bibinfo {year} {1998})},\ \Eprint
  {http://arxiv.org/abs/astro-ph/9712083} {arXiv:astro-ph/9712083} \BibitemShut
  {NoStop}%
\bibitem [{\citenamefont {Son}(1999)}]{Son:1998my}%
  \BibitemOpen
  \bibfield  {author} {\bibinfo {author} {\bibfnamefont {D.~T.}\ \bibnamefont
  {Son}},\ }\href {\doibase 10.1103/PhysRevD.59.063008} {\bibfield  {journal}
  {\bibinfo  {journal} {Phys. Rev. D}\ }\textbf {\bibinfo {volume} {59}},\
  \bibinfo {pages} {063008} (\bibinfo {year} {1999})},\ \Eprint
  {http://arxiv.org/abs/hep-ph/9803412} {arXiv:hep-ph/9803412 [hep-ph]}
  \BibitemShut {NoStop}%
\bibitem [{\citenamefont {Christensson}\ \emph {et~al.}(2001)\citenamefont
  {Christensson}, \citenamefont {Hindmarsh},\ and\ \citenamefont
  {Brandenburg}}]{Christensson:2000sp}%
  \BibitemOpen
  \bibfield  {author} {\bibinfo {author} {\bibfnamefont {M.}~\bibnamefont
  {Christensson}}, \bibinfo {author} {\bibfnamefont {M.}~\bibnamefont
  {Hindmarsh}}, \ and\ \bibinfo {author} {\bibfnamefont {A.}~\bibnamefont
  {Brandenburg}},\ }\href {\doibase 10.1103/PhysRevE.64.056405} {\bibfield
  {journal} {\bibinfo  {journal} {Phys. Rev. E}\ }\textbf {\bibinfo {volume}
  {64}},\ \bibinfo {pages} {056405} (\bibinfo {year} {2001})},\ \Eprint
  {http://arxiv.org/abs/astro-ph/0011321} {arXiv:astro-ph/0011321} \BibitemShut
  {NoStop}%
\bibitem [{\citenamefont {Sigl}(2002)}]{Sigl:2002kt}%
  \BibitemOpen
  \bibfield  {author} {\bibinfo {author} {\bibfnamefont {G.}~\bibnamefont
  {Sigl}},\ }\href {\doibase 10.1103/PhysRevD.66.123002} {\bibfield  {journal}
  {\bibinfo  {journal} {Phys. Rev. D}\ }\textbf {\bibinfo {volume} {66}},\
  \bibinfo {pages} {123002} (\bibinfo {year} {2002})},\ \Eprint
  {http://arxiv.org/abs/astro-ph/0202424} {arXiv:astro-ph/0202424} \BibitemShut
  {NoStop}%
\bibitem [{\citenamefont {Banerjee}\ and\ \citenamefont
  {Jedamzik}(2004)}]{Banerjee:2004df}%
  \BibitemOpen
  \bibfield  {author} {\bibinfo {author} {\bibfnamefont {R.}~\bibnamefont
  {Banerjee}}\ and\ \bibinfo {author} {\bibfnamefont {K.}~\bibnamefont
  {Jedamzik}},\ }\href {\doibase 10.1103/PhysRevD.70.123003} {\bibfield
  {journal} {\bibinfo  {journal} {Phys. Rev. D}\ }\textbf {\bibinfo {volume}
  {70}},\ \bibinfo {pages} {123003} (\bibinfo {year} {2004})},\ \Eprint
  {http://arxiv.org/abs/astro-ph/0410032} {arXiv:astro-ph/0410032} \BibitemShut
  {NoStop}%
\bibitem [{\citenamefont {Campanelli}(2007)}]{Campanelli:2007tc}%
  \BibitemOpen
  \bibfield  {author} {\bibinfo {author} {\bibfnamefont {L.}~\bibnamefont
  {Campanelli}},\ }\href {\doibase 10.1103/PhysRevLett.98.251302} {\bibfield
  {journal} {\bibinfo  {journal} {Phys. Rev. Lett.}\ }\textbf {\bibinfo
  {volume} {98}},\ \bibinfo {pages} {251302} (\bibinfo {year} {2007})},\
  \Eprint {http://arxiv.org/abs/0705.2308} {arXiv:0705.2308 [astro-ph]}
  \BibitemShut {NoStop}%
\bibitem [{\citenamefont {Kahniashvili}\ \emph {et~al.}(2010)\citenamefont
  {Kahniashvili}, \citenamefont {Brandenburg}, \citenamefont {Tevzadze},\ and\
  \citenamefont {Ratra}}]{Kahniashvili:2010gp}%
  \BibitemOpen
  \bibfield  {author} {\bibinfo {author} {\bibfnamefont {T.}~\bibnamefont
  {Kahniashvili}}, \bibinfo {author} {\bibfnamefont {A.}~\bibnamefont
  {Brandenburg}}, \bibinfo {author} {\bibfnamefont {A.~G.}\ \bibnamefont
  {Tevzadze}}, \ and\ \bibinfo {author} {\bibfnamefont {B.}~\bibnamefont
  {Ratra}},\ }\href {\doibase 10.1103/PhysRevD.81.123002} {\bibfield  {journal}
  {\bibinfo  {journal} {Phys. Rev. D}\ }\textbf {\bibinfo {volume} {81}},\
  \bibinfo {pages} {123002} (\bibinfo {year} {2010})},\ \Eprint
  {http://arxiv.org/abs/1004.3084} {arXiv:1004.3084 [astro-ph.CO]} \BibitemShut
  {NoStop}%
\bibitem [{\citenamefont {Grappin}\ \emph {et~al.}(1982)\citenamefont
  {Grappin}, \citenamefont {Frisch}, \citenamefont {Pouquet},\ and\
  \citenamefont {L\'eorat}}]{Grappin82}%
  \BibitemOpen
  \bibfield  {author} {\bibinfo {author} {\bibfnamefont {R.}~\bibnamefont
  {Grappin}}, \bibinfo {author} {\bibfnamefont {U.}~\bibnamefont {Frisch}},
  \bibinfo {author} {\bibfnamefont {A.}~\bibnamefont {Pouquet}}, \ and\
  \bibinfo {author} {\bibfnamefont {J.}~\bibnamefont {L\'eorat}},\ }\href
  {http://adsabs.harvard.edu/abs/1982A%26A...105....6G} {\bibfield  {journal}
  {\bibinfo  {journal} {Astron. Astrophys.}\ }\textbf {\bibinfo {volume}
  {105}},\ \bibinfo {pages} {6} (\bibinfo {year} {1982})}\BibitemShut {NoStop}%
\bibitem [{\citenamefont {Grappin}\ \emph {et~al.}(1983)\citenamefont
  {Grappin}, \citenamefont {Pouquet},\ and\ \citenamefont
  {L\'eorat}}]{Grappin83}%
  \BibitemOpen
  \bibfield  {author} {\bibinfo {author} {\bibfnamefont {R.}~\bibnamefont
  {Grappin}}, \bibinfo {author} {\bibfnamefont {A.}~\bibnamefont {Pouquet}}, \
  and\ \bibinfo {author} {\bibfnamefont {J.}~\bibnamefont {L\'eorat}},\ }\href
  {http://adsabs.harvard.edu/abs/1983A%26A...126...51G} {\bibfield  {journal}
  {\bibinfo  {journal} {Astron. Astrophys.}\ }\textbf {\bibinfo {volume}
  {126}},\ \bibinfo {pages} {51} (\bibinfo {year} {1983})}\BibitemShut
  {NoStop}%
\bibitem [{\citenamefont {Kulsrud}\ and\ \citenamefont
  {Anderson}(1992)}]{Kulsrud:1992rk}%
  \BibitemOpen
  \bibfield  {author} {\bibinfo {author} {\bibfnamefont {R.~M.}\ \bibnamefont
  {Kulsrud}}\ and\ \bibinfo {author} {\bibfnamefont {S.~W.}\ \bibnamefont
  {Anderson}},\ }\href {\doibase 10.1086/171743} {\bibfield  {journal}
  {\bibinfo  {journal} {Astrophys. J.}\ }\textbf {\bibinfo {volume} {396}},\
  \bibinfo {pages} {606} (\bibinfo {year} {1992})}\BibitemShut {NoStop}%
\bibitem [{\citenamefont {Jedamzik}\ and\ \citenamefont
  {Sigl}(2011)}]{PhysRevD.83.103005}%
  \BibitemOpen
  \bibfield  {author} {\bibinfo {author} {\bibfnamefont {K.}~\bibnamefont
  {Jedamzik}}\ and\ \bibinfo {author} {\bibfnamefont {G.}~\bibnamefont
  {Sigl}},\ }\href {\doibase 10.1103/PhysRevD.83.103005} {\bibfield  {journal}
  {\bibinfo  {journal} {Phys. Rev. D}\ }\textbf {\bibinfo {volume} {83}},\
  \bibinfo {pages} {103005} (\bibinfo {year} {2011})},\ \Eprint
  {http://arxiv.org/abs/1012.4794} {arXiv:1012.4794 [astro-ph.CO]} \BibitemShut
  {NoStop}%
\bibitem [{rem({\natexlab{a}})}]{remark1}%
  \BibitemOpen
  \href@noop {} {} \bibinfo {note} {{We clarify here that the 
  assumption of an initial white noise velocity spectrum is not essential for the results
  obtained in \cite{PhysRevD.83.103005}}}\BibitemShut {Stop}%
\bibitem [{rem({\natexlab{b}})}]{remark2}%
  \BibitemOpen
  \href@noop {} {} \bibinfo {note} {{Note that energy is
  trivially conserved to first order in $\Delta t$.}}\BibitemShut {Stop}%
\bibitem [{\citenamefont {Batchelor}\ and\ \citenamefont
  {Proudman}(1956)}]{Batchelor1956}%
  \BibitemOpen
  \bibfield  {author} {\bibinfo {author} {\bibfnamefont {G.~K.}\ \bibnamefont
  {Batchelor}}\ and\ \bibinfo {author} {\bibfnamefont {I.}~\bibnamefont
  {Proudman}},\ }\href {http://www.jstor.org/stable/91587} {\bibfield
  {journal} {\bibinfo  {journal} {Phil. Trans. R. Soc. London A}\ }\textbf
  {\bibinfo {volume} {248}},\ \bibinfo {pages} {369} (\bibinfo {year}
  {1956})}\BibitemShut {NoStop}%
\bibitem [{\citenamefont {Durrer}\ and\ \citenamefont
  {Caprini}(2003)}]{Durrer:2003ja}%
  \BibitemOpen
  \bibfield  {author} {\bibinfo {author} {\bibfnamefont {R.}~\bibnamefont
  {Durrer}}\ and\ \bibinfo {author} {\bibfnamefont {C.}~\bibnamefont
  {Caprini}},\ }\href {\doibase 10.1088/1475-7516/2003/11/010} {\bibfield
  {journal} {\bibinfo  {journal} {J. Cosmol. Astropart. Phys.}\ }\textbf {\bibinfo {volume} {11}},\
  \bibinfo {pages} {010} (\bibinfo {year} {2003})},\ \Eprint
  {http://arxiv.org/abs/astro-ph/0305059} {arXiv:astro-ph/0305059} \BibitemShut
  {NoStop}%
\bibitem [{\citenamefont {Tevzadze}\ \emph {et~al.}(2012)\citenamefont
  {Tevzadze}, \citenamefont {Kisslinger}, \citenamefont {Brandenburg},\ and\
  \citenamefont {Kahniashvili}}]{Tevzadze:2012kk}%
  \BibitemOpen
  \bibfield  {author} {\bibinfo {author} {\bibfnamefont {A.~G.}\ \bibnamefont
  {Tevzadze}}, \bibinfo {author} {\bibfnamefont {L.}~\bibnamefont
  {Kisslinger}}, \bibinfo {author} {\bibfnamefont {A.}~\bibnamefont
  {Brandenburg}}, \ and\ \bibinfo {author} {\bibfnamefont {T.}~\bibnamefont
  {Kahniashvili}}\ }\href@noop {} {({2012})},\ \href {\doibase 10.1088/0004-637X/759/1/54 } {\bibfield
  {journal} {\bibinfo  {journal} {Astrophys. J.}\ }\textbf {\bibinfo {volume} {759}},\
  \bibinfo {pages} {54} (\bibinfo {year} {2012})},\  \Eprint
  {http://arxiv.org/abs/1207.0751} {arXiv:1207.0751 [astro-ph.CO]} \BibitemShut
  {NoStop}%
\bibitem [{\citenamefont {von~K\'{a}rm\'{a}n}\ and\ \citenamefont
  {Howarth}(1938)}]{ProcRSocLondonA.164.192}%
  \BibitemOpen
  \bibfield  {author} {\bibinfo {author} {\bibfnamefont {T.}~\bibnamefont
  {von~K\'{a}rm\'{a}n}}\ and\ \bibinfo {author} {\bibfnamefont {L.}~\bibnamefont
  {Howarth}},\ }\href {http://www.jstor.org/discover/10.2307/97087} {\bibfield
  {journal} {\bibinfo  {journal} {Proc. R. Soc. London A}\ }\textbf {\bibinfo
  {volume} {164}},\ \bibinfo {pages} {192} (\bibinfo {year}
  {1938})}\BibitemShut {NoStop}%
\end{thebibliography}
\end{document}